\documentclass[11pt]{article}
\usepackage{acl}
\usepackage{times}
\usepackage{latexsym}
\usepackage[T1]{fontenc}
\usepackage[utf8]{inputenc}
\usepackage{microtype}
\usepackage{inconsolata}
\usepackage{graphicx}
\usepackage{amsmath} 
\usepackage{amsfonts}
\usepackage{amssymb}
\usepackage{booktabs}
\usepackage{lipsum}
\usepackage{multirow} 

\title{ReStyle-TTS: Relative and Continuous Style Control for Zero-Shot Speech Synthesis}
\author{
Haitao Li$^{1,2}$, Chunxiang Jin$^{3}$, Chenglin Li$^{1,2}$, Wenhao Guan$^{4,2}$, Zhengxing Huang$^{1}$, Xie Chen$^{5,2}$ \\
$^{1}$Zhejiang University,
$^{2}$Shanghai Innovation Institute,
$^{3}$Ant Group, \\
$^{4}$Xiamen University,
$^{5}$Shanghai Jiao Tong University \\
\texttt{lihaitao@zju.edu.cn, chenxie95@sjtu.edu.cn}
}

\begin{document}
\maketitle
\begin{abstract}
Zero-shot text-to-speech models can clone a speaker’s timbre from a short reference audio, but they also strongly inherit the speaking style present in the reference. As a result, synthesizing speech with a desired style often requires carefully selecting reference audio, which is impractical when only limited or mismatched references are available. While recent controllable TTS methods attempt to address this issue, they typically rely on absolute style targets and discrete textual prompts, and therefore do not support continuous and reference-relative style control.
We propose ReStyle-TTS, a framework that enables continuous and reference-relative style control in zero-shot TTS. Our key insight is that effective style control requires first reducing the model’s implicit dependence on reference style before introducing explicit control mechanisms. To this end, we introduce Decoupled Classifier-Free Guidance (DCFG), which independently controls text and reference guidance, reducing reliance on reference style while preserving text fidelity. On top of this, we apply style-specific LoRAs together with Orthogonal LoRA Fusion to enable continuous and disentangled multi-attribute control, and introduce a Timbre Consistency Optimization module to mitigate timbre drift caused by weakened reference guidance.
Experiments show that ReStyle-TTS enables user-friendly, continuous, and relative control over pitch, energy, and multiple emotions while maintaining intelligibility and speaker timbre, and performs robustly in challenging mismatched reference–target style scenarios.
The project webpage is available at https://cucl-2.github.io/Restyle-TTS.
\end{abstract}

\begin{table*}[t]
\centering
\small
\setlength{\tabcolsep}{8pt}
\begin{tabular}{lcccc}
\toprule
\textbf{Model} & \textbf{Timbre Source} & \textbf{Style Source} & \textbf{Continuous Control} & \textbf{Control Type} \\
\midrule
IndexTTS2 / Vevo & Reference Audio & Style Audio & No & Absolute \\
ControlSpeech / EmoVoice / CosyVoice & Reference Audio & Text Description & No & Absolute \\
StyleFusion TTS & Reference Audio & Audio or Text & No & Absolute \\
\textbf{ReStyle-TTS (Ours)} & Reference Audio & Style LoRA & \textbf{Yes} & \textbf{Relative} \\
\bottomrule
\end{tabular}
\caption{Comparison of controllable zero-shot TTS methods.}
\label{tab:controllable_zero_shot_tts}
\end{table*}

\section{Introduction}
Recent zero-shot text-to-speech (TTS) systems can synthesize speech for unseen speakers from only a short reference audio clip. By conditioning on this reference, these models can preserve the speaker’s identity (timbre) while following the input text. However, the generated speech is often strongly influenced by the speaking style present in the reference audio, including prosody and emotion, which fundamentally limits controllability in zero-shot TTS. As a result, synthesizing speech with a desired style often requires carefully selecting reference audio that matches the target style, which is time-consuming and sometimes impossible when only limited or mismatched reference audio is available. This issue is especially pronounced when the available reference conveys a different style from the target, such as attempting to generate angry speech when only a happy reference clip is available.

This limitation naturally motivates research on controllable TTS like InstructTTS \citep{guo2023prompttts,yang2024instructtts,liu2023promptstyle}. While these approaches have demonstrated promising results, most of them assume a fixed or predefined speaker space and therefore do not support true zero-shot speaker generalization from short reference audio. More recent work attempts to bridge the gap between voice cloning and controllability by enabling timbre cloning while allowing style manipulation. IndexTTS2 \citep{zhou2025indextts2} and Vevo \citep{zhang2025vevo} achieves style control through a style prompt audio, but still requires carefully selecting suitable reference samples. ControlSpeech \citep{ji2024controlspeech}, EmoVoice \citep{yang2025emovoice}, and CosyVoice \citep{du2024cosyvoice2,du2025cosyvoice} instead allow users to specify style through language-based prompts on top of voice cloning, which is more user-friendly. However, text-based style control remains unstable due to the complex and many-to-many relationship between textual descriptions and acoustic realizations. Moreover, these methods typically rely on absolute target styles and do not support continuous and reference-relative style control, where attributes are adjusted incrementally with respect to the reference, which is more intuitive and user-friendly. A structured comparison of representative controllable zero-shot TTS methods is summarized in Table~\ref{tab:controllable_zero_shot_tts}.

Achieving continuous and reference-relative style control while preserving zero-shot voice cloning capability is challenging due to a fundamental trade-off. If the model remains strongly dependent on the reference audio, the generated speech is tightly constrained by the reference style, leaving little room for flexible control. Conversely, simply weakening the influence of the reference audio often degrades speaker timbre consistency, undermining the core objective of zero-shot TTS. To address this challenge, we propose ReStyle-TTS. We first introduce Decoupled Classifier-Free Guidance (DCFG), which separately controls the guidance strengths from text and from the reference audio, allowing us to reduce the reliance on the reference audio during generation while maintaining text fidelity. Building on this, we apply style-specific LoRAs together with Orthogonal LoRA Fusion to inject explicit and continuously adjustable style factors (e.g., pitch, energy, and emotions) on top of the base model. Orthogonal LoRA Fusion enables the independent and simultaneous control of multiple style attributes. Finally, since reducing reference guidance can introduce timbre drift, we incorporate a Timbre Consistency Optimization module that explicitly reinforces speaker timbre preservation during training.

With these components, ReStyle-TTS enables controllable zero-shot TTS that provides user-friendly, continuous, and relative control over speaking style while preserving speaker timbre. We also evaluate our method on several challenging scenarios, including generating angry speech from happy references, a case that previous approaches have not effectively addressed.

Our contributions are summarized as follows:
\begin{itemize}
    \item We propose \textbf{ReStyle-TTS}, a controllable zero-shot TTS framework that enables user-friendly, continuous, and reference-relative control of speaking style while preserving speaker timbre.
    \item We demonstrate that ReStyle-TTS can effectively control pitch, energy, and emotions, and it excels in handling scenarios where the reference and target styles are mismatched.
\end{itemize}

\section{Related Works}
\noindent \textbf{Zero-shot TTS.}  
Zero-shot text-to-speech (TTS) aims to synthesize speech for unseen speakers without explicit speaker-specific training and can be broadly categorized into non-autoregressive (NAR), autoregressive (AR), and hybrid architectures. 
In the NAR domain, Voicebox \citep{le2023voicebox} formulates TTS as a text-guided speech infilling problem, trained via flow matching \citep{lipman2022flow}. E2-TTS \citep{eskimez2024e2} and F5-TTS \citep{chen2024f5} simplify the alignment process by appending filler tokens to the text sequence, avoiding the need for duration models.
In AR-based approaches, models such as AudioLM \citep{borsos2023audiolm}, VALL-E \citep{wang2023neural}, and SparkTTS \citep{wang2025spark} model discrete audio semantic and acoustic tokens, leveraging powerful language modeling techniques for speech generation. 
Hybrid architectures like CosyVoice \citep{du2024cosyvoice, du2024cosyvoice2, du2025cosyvoice}, Seed-TTS \citep{anastassiou2024seed}, and IndexTTS2 \citep{zhou2025indextts2} autoregressively model semantic tokens and then employ flow matching to generate mel spectrograms. To reduce the information loss caused by discrete token modeling, continuous token modeling has been explored in DiTAR \citep{jia2025ditar} and MELLE \citep{meng2024autoregressive}, inspired by continuous representation learning in image generation, such as MAR \citep{li2024autoregressive}.

All these models perform well in zero-shot TTS, but they often inherit the speaking style of the reference. This makes synthesizing speech in a desired style time-consuming, as it requires carefully selecting reference audio, which may be infeasible when suitable references are unavailable. Ideally, synthesized speech should allow for flexible style control.

\noindent \textbf{Controllable Speech Synthesis.}  
Early controllable TTS models, such as FastSpeech2 \citep{ren2020fastspeech} and FastPitch \citep{lancucki2021fastpitch}, primarily controlled prosody by explicitly predicting low-level attributes like pitch, energy, and duration. Later advancements introduced control through discrete textual tags \citep{kim2021expressive,gao2025emo} and moved toward prompt-based control, where natural language descriptions specify the desired speaking style. Notable models include InstructTTS \citep{yang2024instructtts}, PromptStyle \citep{liu2023promptstyle}, and PromptTTS \citep{guo2023prompttts}. However, these models are either speaker-independent or rely on predefined speaker identities or embeddings for timbre, meaning they cannot perform true zero-shot speaker cloning from a brief reference audio clip.

More recent approaches attempt to bridge the gap between voice cloning and controllability by enabling timbre cloning while allowing style manipulation. SC VALL-E \citep{kim2023sc} achieves style control by adjusting a latent style control vector; however, this vector itself is not interpretable. Vevo \citep{zhang2025vevo} and IndexTTS2 \citep{zhou2025indextts2} achieve style control by providing a separate style prompt audio. ControlSpeech \citep{ji2024controlspeech}, EmoVoice \citep{yang2025emovoice}, and CosyVoice \citep{du2024cosyvoice2,du2025cosyvoice} instead enable style control through language-based style prompts on top of voice cloning, which is more user-friendly. StyleFusion TTS \citep{chen2024stylefusion} further supports style control using both textual descriptions and style audio simultaneously.

However, all these methods cannot support continuous or relative control, and they disregard the inherent style of the reference audio. A more user-friendly interaction would instead allow relative adjustments, such as slightly increasing the pitch or making the speech sound a bit angrier.

\noindent \textbf{Style Control in Image Generation using LoRA.} In the field of image generation, it is common practice to train LoRA models to modify the style of generated images \citep{gandikota2024concept,frenkel2024implicit} and to combine multiple LoRA models for controlling a blend of styles \citep{shah2024ziplora,zhong2024multi,ouyang2025k,zheng2025freelora}. However, in the TTS domain, the style of speech is inherently embedded in the reference audio. The model is trained to replicate the style from the reference audio to the generated audio. As a result, the direct application of LoRA-based style control methods from image generation is not suitable for TTS.

\begin{figure*}[t]
    \centering
    \includegraphics[width=0.90\linewidth]{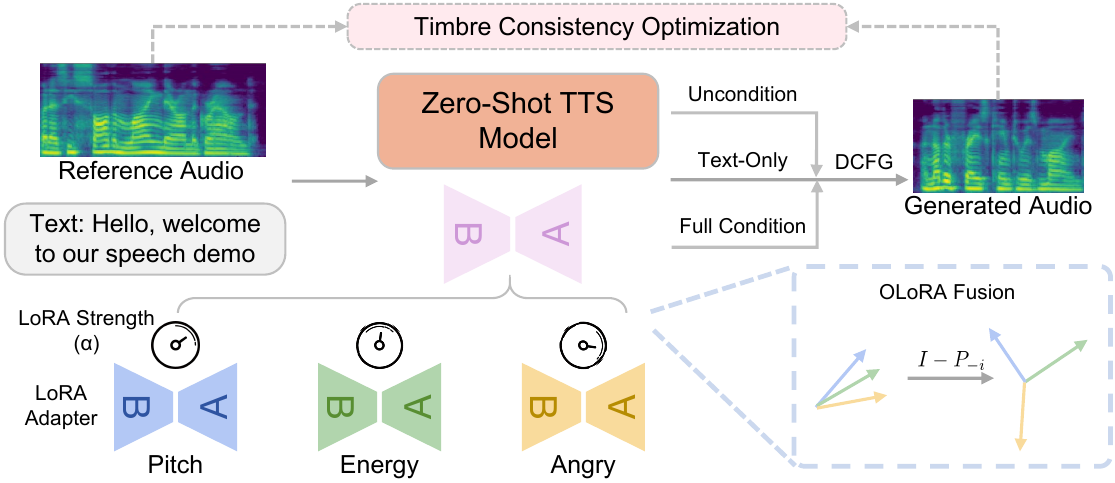}
    \caption{The overall framework of \textbf{ReStyle-TTS}. The proposed method consists of three logically coordinated components: (1) \textbf{Decoupled Classifier-Free Guidance (DCFG)} reduces the model's dependency on the reference style while maintaining text fidelity; (2) \textbf{Orthogonal LoRA Fusion (OLoRA)} reduces interference among multiple style-specific LoRAs by projecting each LoRA onto the orthogonal complement of the subspace spanned by the others; and (3) \textbf{Timbre Consistency Optimization (TCO)} reinforces speaker identity preservation through a similarity-based reward mechanism.}
    \label{fig:overview}
\end{figure*}

\section{Method}
\subsection{Overview}
Style control using LoRA has been widely adopted in image generation, where LoRAs fine-tuned on specific datasets (e.g., anime or Van Gogh paintings) can control image styles during inference. However, this approach doesn’t directly apply to zero-shot TTS systems due to the key difference in the use of reference audio. In image generation, outputs are guided solely by text prompts, while zero-shot TTS systems rely on both text and reference audio during inference. The model learns to replicate not only the timbre but also the style of the reference audio, making precise style control difficult.

To address this issue, we propose ReStyle-TTS, which uses Decoupled Classifier-Free Guidance (DCFG) to reduce the model’s dependency on the reference audio while maintaining faithfulness to the text. After this decoupling step, a series of Style LoRA modules can be trained to control various attributes such as pitch or emotion, and we further introduce an Orthogonal LoRA Fusion mechanism to combine multiple Style LoRAs without mutual interference. However, weakening the reference dependency may reduce timbre consistency. Therefore, we additionally propose a Timbre Consistency Optimization (TCO) module to explicitly preserve the speaker timbre encoded in the reference audio. An overview of ReStyle-TTS is shown in Figure \ref{fig:overview}.

\subsection{Decoupled Classifier-Free Guidance}
In standard zero-shot TTS systems, classifier-free guidance (CFG) is commonly used to balance conditional and unconditional predictions during generation \citep{du2024cosyvoice,du2024cosyvoice2,chen2024f5}. Let $f_{a,t}$ denote the predicted audio representation conditioned on both the reference audio $a$ and the text $t$, and let $f_{\varnothing,\varnothing}$ denote the unconditional prediction. The conventional CFG formulation is:
\begin{align}
\hat{f} = f_{a,t} + \lambda_{\text{cfg}} (f_{a,t} - f_{\varnothing,\varnothing}),
\end{align}
where $\lambda_{\text{cfg}}$ controls the overall guidance strength. Increasing $\lambda_{\text{cfg}}$ enhances the influence of the conditional inputs but does not distinguish between text guidance and reference guidance, since both are entangled within $f_{a,t}$. As a result, adjusting $\lambda_{\text{cfg}}$ simultaneously affects text fidelity and style dependency—reducing the weight of the reference also weakens textual alignment.

To disentangle these effects, we introduce Decoupled Classifier-Free Guidance. We separately compute intermediate predictions conditioned on (i) text only, $f_{\varnothing,t}$, and (ii) both reference and text, $f_{a,t}$. DCFG combines them as follows:
\begin{align}
\hat{f}_{\text{DCFG}}
= f_{\varnothing,t}
+ \lambda_t (f_{\varnothing,t} - f_{\varnothing,\varnothing})
+ \lambda_a (f_{a,t} - f_{\varnothing,t}),
\end{align}
where $\lambda_t$ and $\lambda_a$ are independent guidance strengths for text and reference. Specifically, $\lambda_t$ controls how strongly the model follows the text, while $\lambda_a$ determines how much it depends on the reference audio.  

When $\lambda_t = \lambda_{\text{cfg}}$ and $\lambda_a = 1 + \lambda_{\text{cfg}}$, DCFG reduces to the standard CFG formulation. By lowering $\lambda_a$ while keeping $\lambda_t$ fixed, we explicitly reduce the model’s reliance on the reference style without harming text alignment. This makes subsequent style control feasible, rather than relying entirely on the style present in the reference audio.

\subsection{Style LoRA and Orthogonal LoRA Fusion}
With DCFG reducing the model’s dependency on reference style, the generated audio is no longer bound to the prosody or emotion of the reference. This enables us to introduce controllable style modification using LoRA, inspired by its successful application in image generation \citep{zhong2024multi,ouyang2025k,zheng2025freelora}. Similarly, we fine-tune style-specific LoRAs on audio datasets annotated with particular attributes such as high/low pitch or different emotions. Each LoRA thus captures a single interpretable attribute direction in the model parameter space.

Following the practice in the image domain, the influence of each LoRA can be continuously adjusted by scaling its magnitude, enabling smooth control of style intensity \citep{gandikota2024concept}. Moreover, since each LoRA specializes in a single attribute, it is desirable to combine multiple LoRAs to control several attributes simultaneously. However, directly adding LoRA weights often leads to interference between adapters \citep{shah2024ziplora,zhong2024multi,ouyang2025k,zheng2025freelora}, resulting in entangled or unstable styles.

To address this, we propose Orthogonal LoRA Fusion (OLoRA), a training-free mechanism for combining multiple style LoRAs. OLoRA jointly orthogonalizes the parameter subspaces of individual LoRAs and performs weighted fusion to compose multiple style attributes without retraining. Formally, for a linear layer with $N$ trained LoRAs, let $\{\Delta W_i\}_{i=1}^{N}$ denote their low-rank updates, where $\Delta W_i = B_i A_i$. We first decorrelate them by projecting each $\Delta W_i$ onto the orthogonal complement of the subspace spanned by all others. Denote $v_i = \mathrm{vec}(\Delta W_i) \in \mathbb{R}^{D}$ and $V_{-i} = [v_1, \ldots, v_{i-1}, v_{i+1}, \ldots, v_N]$, and compute the projection matrix $P_{-i} = V_{-i} (V_{-i})^{+}$ using least squares or SVD. The orthogonalized update is then $\tilde{v}_i = (I - P_{-i}) v_i$, which is reshaped back to $\tilde{\Delta W}_i$. In contrast to sequential projection schemes that are sensitive to the fusion order and may yield inconsistent compositions, OLoRA performs joint orthogonalization by projecting each adapter against the entire subspace spanned by the remaining ones, resulting in an order-independent fusion process. Crucially, since the number of LoRAs is significantly smaller than the parameter dimension ($N \ll D$), the style vectors occupy a sparse subspace within the high-dimensional manifold. This sparsity ensures that orthogonal projection effectively eliminates interference.

The orthogonalized LoRAs are fused through a weighted combination, and the final generation of ReStyle-TTS is expressed using the following unified formulation:
\begin{gather}
\hat{f}_{\text{ReStyle}} = g_{\varnothing,t}
+ \lambda_t\big(g_{\varnothing,t} - g_{\varnothing,\varnothing}\big)
+ \lambda_a\big(g_{a,t} - g_{\varnothing,t}\big), \\
g_{a,t} = f_{a,t}^{(\Theta + \Delta W_{\text{fuse}})}, \\
\Delta W_{\text{fuse}} = \sum_{i=1}^{N} \alpha_i \tilde{\Delta W}_i.
\end{gather}
Here, $\Theta$ denotes the base model parameters, $\lambda_t$ and $\lambda_a$ control the text and reference guidance strengths, and each $\alpha_i$ provides continuous control over its corresponding style attribute.

\subsection{Timbre Consistency Optimization}
While DCFG relaxes the dependency on the reference audio and OLoRA enables flexible style control, these modifications may weaken the preservation of speaker timbre. To explicitly enhance timbre consistency without altering the main training objective, we introduce Timbre Consistency Optimization (TCO), a lightweight reinforcement strategy guided by speaker similarity rewards.

In standard flow-matching training, the model parameters \(\theta\) are optimized by minimizing the mean squared error between the predicted and target flows: \(\mathcal{L}_{\text{FM}}(\theta) = \mathbb{E}_{(x, y)} \| f_\theta(x) - y \|_2^2\). To incorporate timbre feedback, we sample speech generated by the current model and evaluate its speaker similarity against the corresponding reference audio, which serves as a reward signal \(r\). To reduce reward variance, we maintain an exponential moving average (EMA) baseline \(b_t = \mu b_{t-1} + (1-\mu) r_t\), and define the advantage as \(A_t = r_t - b_t\). To avoid the training instability and computational overhead of policy gradients, we instead adopt an advantage-weighted regression strategy \citep{peng2019advantage} that reweights the flow-matching loss using a smooth, bounded weight \(w_t = 1 + \lambda \tanh(\beta A_t)\), where \(\lambda\) controls the reward strength and \(\beta\) modulates sensitivity to advantage. The total objective becomes \(\mathcal{L}_{\text{total}} = w_t \cdot \mathcal{L}_{\text{FM}}\).

This formulation can be viewed as a reward-modulated weighting of the original supervised loss. Samples with higher speaker similarity receive stronger gradient emphasis, while those with lower similarity are naturally down-weighted. Since no gradient is propagated through the generation or reward computation, TCO preserves the stability and efficiency of standard flow-matching training. As a result, TCO effectively reinforces timbre consistency between generated and reference speech.

\begin{figure*}[t]
    \centering
    \includegraphics[width=1\linewidth]{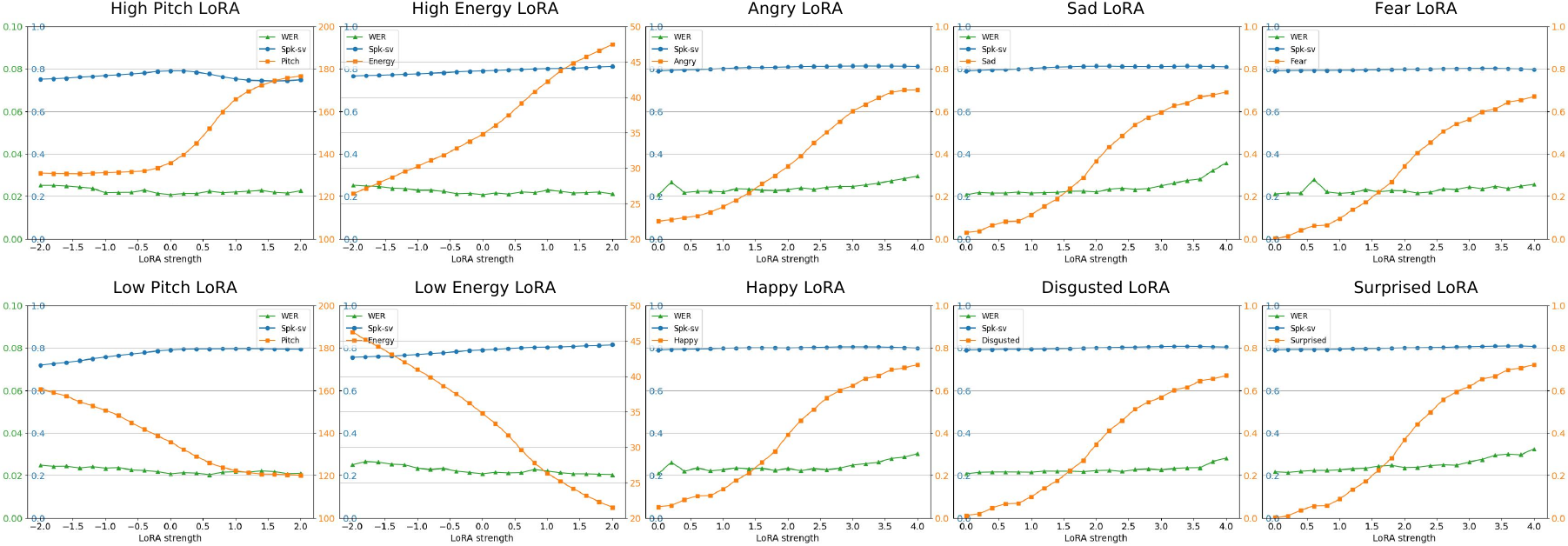}
    \caption{
    Continuous single-attribute control with style-specific LoRAs.}
    \label{fig:single-attr}
\end{figure*}

\section{Experiments}
\subsection{Experimental Setup}
\noindent\textbf{Dataset.} We trained separate LoRA modules on different subsets of the VccmDataset~\citep{ji2024controlspeech}. The VccmDataset is composed of LibriTTS~\citep{zen2019libritts} and several emotion-focused audio datasets~\citep{christophe2016cstr, zhou2021seen, dupuis2010toronto}. Specifically, we used subsets corresponding to high and low pitch, high and low energy, and multiple emotion categories including \textit{angry}, \textit{disgusted}, \textit{fear}, \textit{happy}, \textit{sad}, \textit{surprised}, and \textit{neutral}, while excluding \textit{contempt} due to insufficient data. For evaluation, we conducted controllable zero-shot speech synthesis experiments on the Seed-TTS test set~\citep{anastassiou2024seed} and additionally used the VccmDataset~\citep{ji2024controlspeech} test set for the contradictory-style setting, which requires emotional audio to create mismatched reference–target conditions.

\noindent\textbf{Implementation.} Instead of training the TTS model from scratch, we fine-tune the well-known F5-TTS~\citep{chen2024f5}. During LoRA training, we inject LoRA adapters into all linear layers with a rank of 32 and an alpha value of 64. The AdamW optimizer is used with a learning rate of $1\times10^{-5}$ and a batch size of 30,000 audio frames. Because the amount of audio data varies across subsets, we fixed the total training time to 250 hours rather than keeping the number of epochs constant. In DCFG training, the masked speech input is first dropped with a rate of 0.3, and then the input containing both masked speech and text is dropped with a rate of 0.2. In Timbre Consistency Optimization, the reward strength coefficient is set to $\lambda = 0.2$, the advantage sensitivity to $\beta = 5.0$, and the EMA momentum for the baseline to $\mu = 0.9$. For standard CFG, a common choice is $\lambda_{\text{cfg}} = 2$. When using our DCFG, the setting $\lambda_{t} = 2$ and $\lambda_{a} = 3$ is equivalent to this conventional configuration. In order to reduce the model's dependence on the reference audio, we instead set $\lambda_{a} = 0.5$.

\noindent\textbf{Evaluation.} Following ControlSpeech~\citep{ji2024controlspeech}, we report not only the Word Error Rate (WER) and timbre similarity (Spk-sv) between the reference and synthesized speech, but also measure attribute-specific control effectiveness. For subjective evaluations, we conduct MOS-SA (Mean Opinion Score–Style Accuracy) evaluations to measure the accuracy of the synthesized speech’s style. The evaluation details can be found in Appendix \ref{appendix:eval}.

\subsection{Single-Attribute Control}
To verify that ReStyle-TTS can continuously control individual attributes without harming intelligibility or speaker identity, we first activate a single style LoRA at a time and sweep its strength over a range of values. Figure~\ref{fig:single-attr} summarizes the results for pitch (high/low), energy (high/low), and six emotions (angry, sad, fear, happy, disgusted, surprised). The reported metrics are averaged over the Seed-TTS test set.

For the prosodic LoRAs (pitch and energy), the attribute curves vary smoothly as the LoRA strength changes, while WER and Spk-sv remain almost constant. Notably, negative scaling of a `high-attribute' LoRA naturally produces the opposite effect, effectively enabling bidirectional control even when only one side of the attribute was trained. 
For emotional LoRAs, we similarly obtain monotonic control over the emotion similarity score as the LoRA strength increases. Unlike text-prompt-based methods, where emotion is typically specified by discrete labels or natural language descriptions and is therefore difficult to adjust continuously, our method yields a smooth intensity knob for each emotion. These results confirm that ReStyle-TTS enables precise and continuous single-attribute manipulation for both low-level prosody and high-level emotion.

\subsection{Multi-Attribute Composition}
\begin{figure}[t]
    \centering
    \includegraphics[width=1\linewidth]{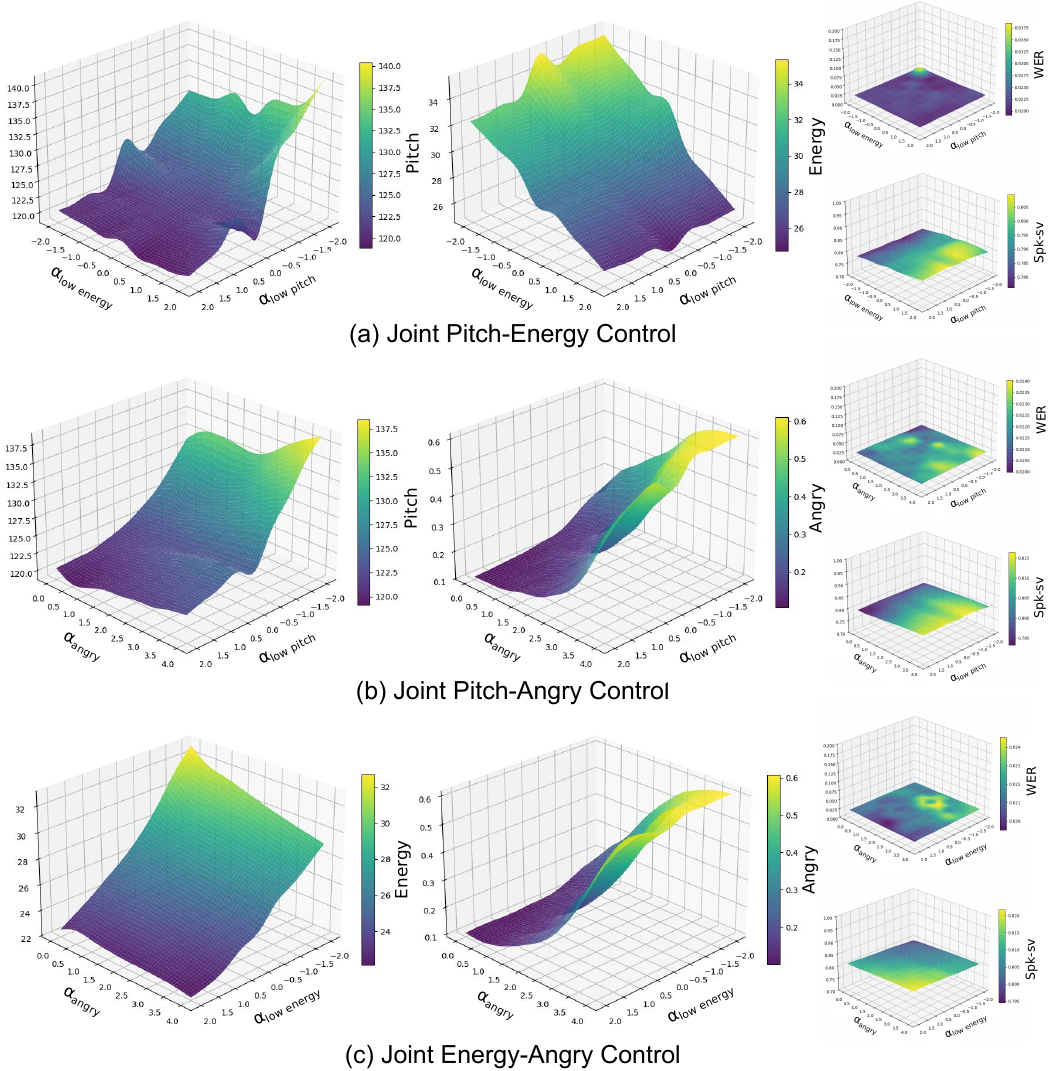}
    \caption{
    Two-attribute joint control results.
    }
    \label{fig:multi-attr}
\end{figure}

To further evaluate whether different Style LoRAs can be jointly applied without introducing noticeable interference, we activate two LoRAs simultaneously and sweep their strengths over a 2D grid. Figure~\ref{fig:multi-attr} presents representative combinations. The reported metrics are averaged over the Seed-TTS test set. Across all evaluated pairs, the controlled attributes vary smoothly along their respective axes. Modulating the strength of one LoRA primarily influences its target attribute, while the other attribute remains largely stable. Meanwhile, both WER and speaker similarity remain stable over the entire 2D space, suggesting that simultaneous multi-attribute manipulation does not compromise intelligibility or timbre preservation.

\begin{figure}[t]
    \centering
    \includegraphics[width=1\linewidth]{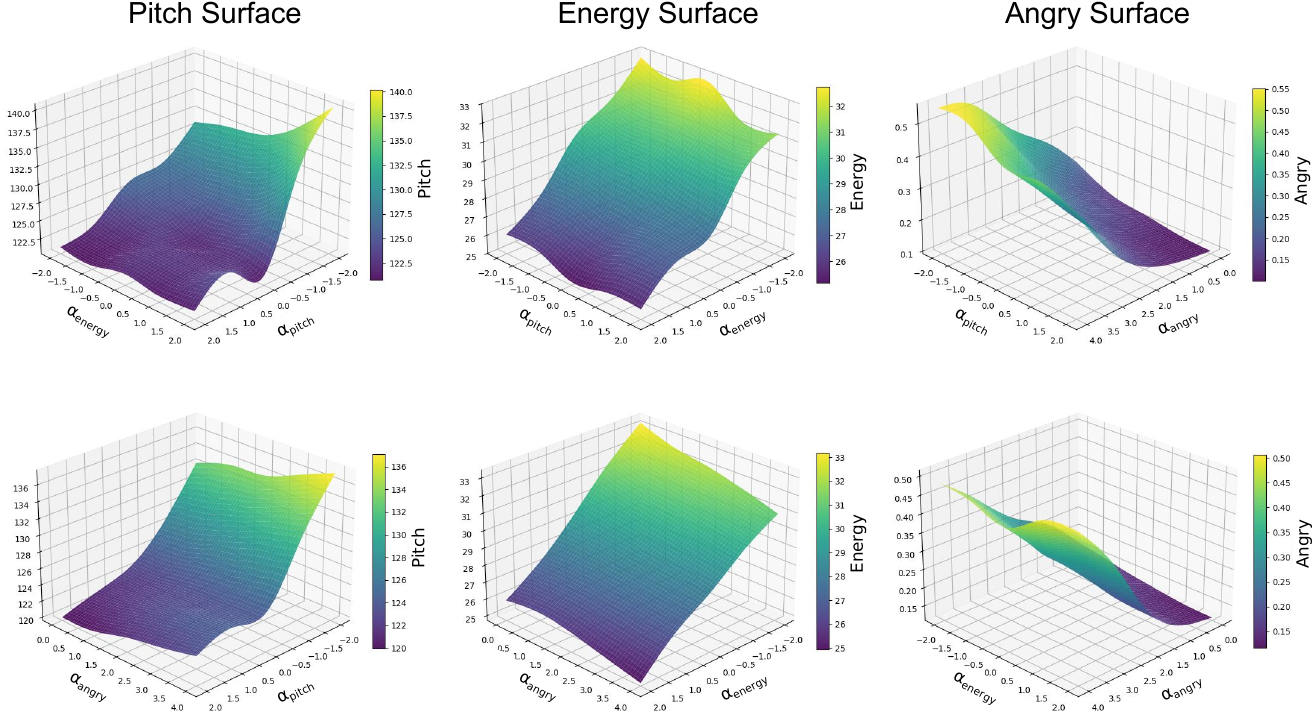}
    \caption{
    Three-attribute joint control results.
    }
    \label{fig:three-attr}
\end{figure}

To further push the analysis, we activate three Style LoRAs simultaneously and evaluate how the model behaves in the resulting three-dimensional control space. As shown in Figure~\ref{fig:three-attr}, the surfaces for pitch, energy, and anger each show smooth and monotonic variation along their respective control axes.

\subsection{Relative Style Control}
\begin{figure}[t]
    \centering
    \includegraphics[width=1\linewidth]{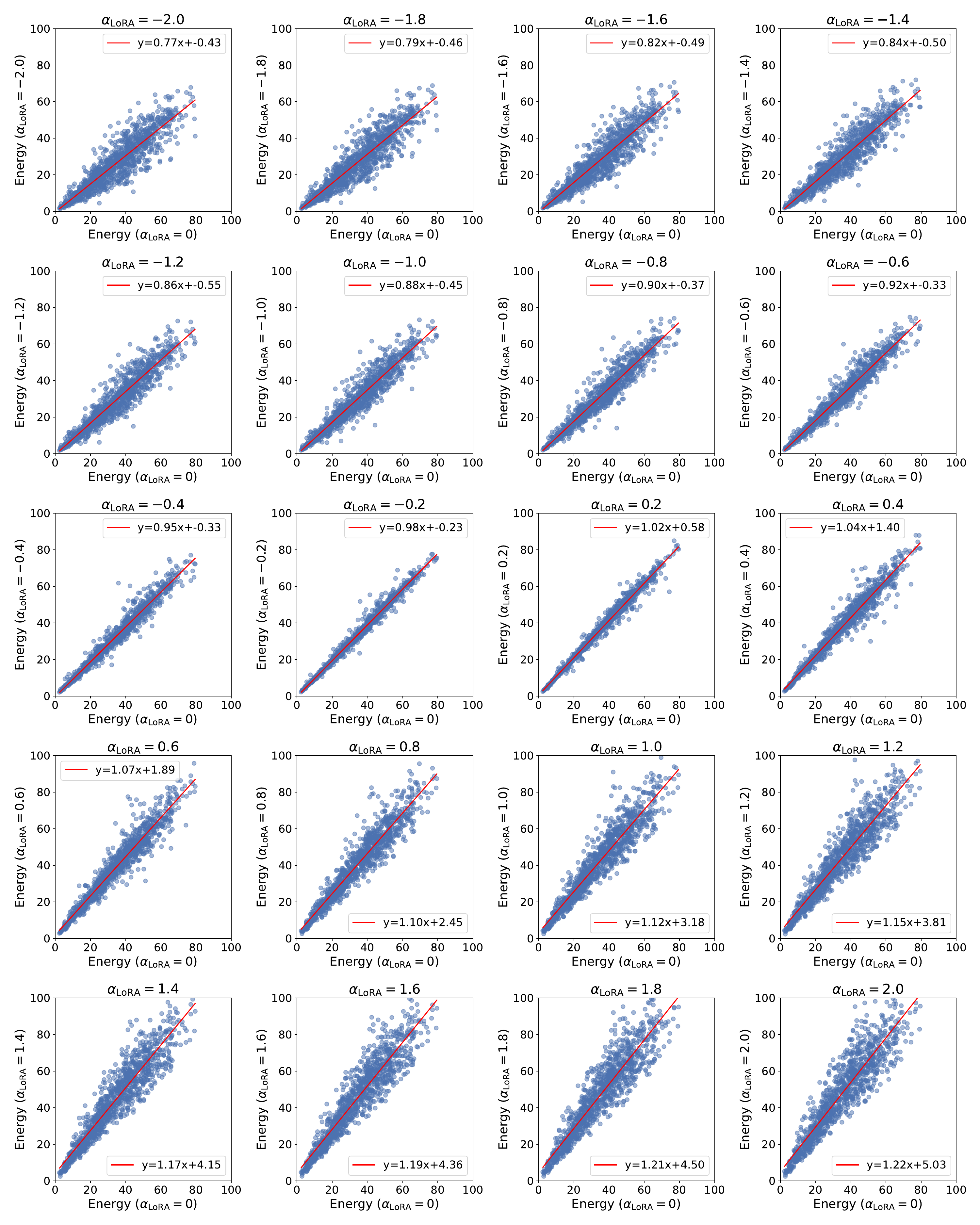}
    \caption{
    Reference energy vs.\ generated energy under different LoRA strengths. 
    }
    \label{fig:relative_high_energy}
\end{figure}

\begin{figure}[t]
    \centering
    \includegraphics[width=0.8\linewidth]{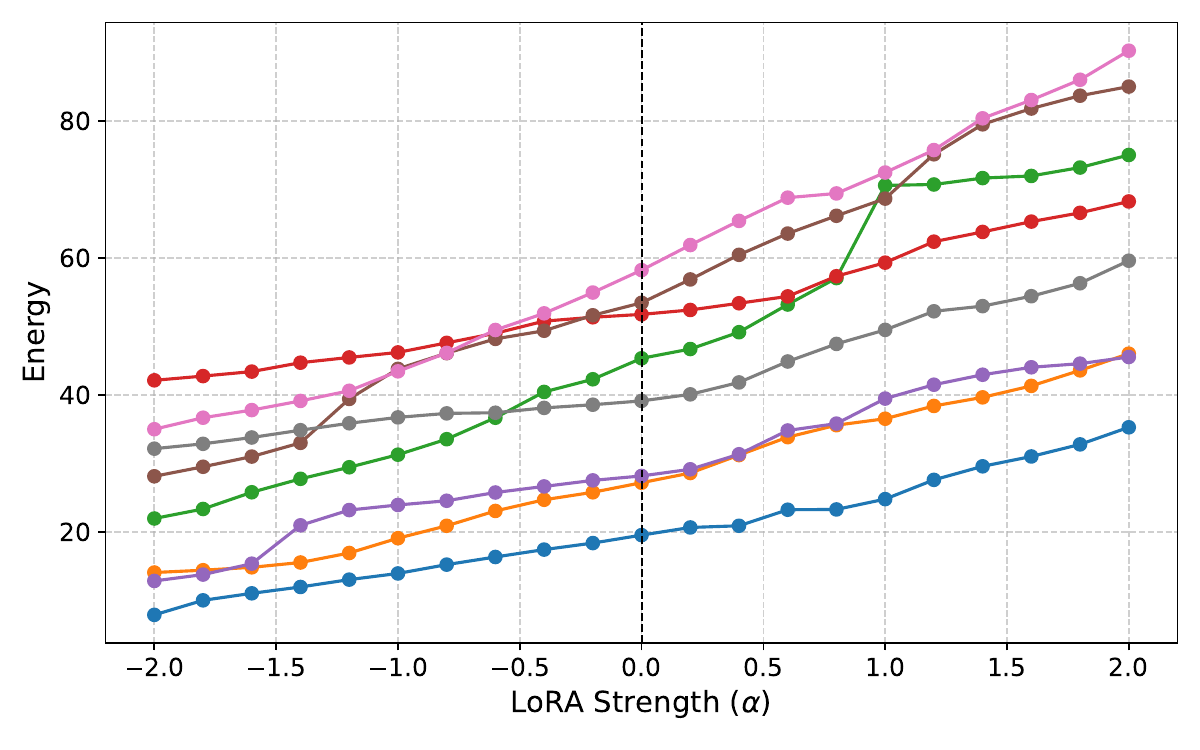}
    \caption{
    Energy trajectories of five randomly selected samples as the High Energy LoRA scale varies.
    }
    \label{fig:relative_high_energy_single}
\end{figure}

A key advantage of ReStyle-TTS is its ability to perform relative style control: attributes are adjusted with respect to the reference audio rather than pushed toward a fixed absolute target. This interaction paradigm is more predictable and user-friendly. Previously, we reported only averages over the entire Seed-TTS test set. To assess relative control, we examine how the Style LoRA affects the attribute of each individual sample. Figure~\ref{fig:relative_high_energy} plots reference energy against generated energy under different LoRA scales. Across all scales, the points form clear linear trends with regression slopes ranging from 0.77 to 1.22 and intercepts consistently near zero. This pattern indicates that the LoRA induces a roughly proportional change. As a result, the relative ordering among reference samples is preserved, in contrast to absolute control, which would drive the slope toward 0 and collapse all samples toward the same target value. Figure~\ref{fig:relative_high_energy_single} shows energy trajectories for eight reference samples. All curves vary smoothly and monotonically while starting from distinct baselines corresponding to each sample’s inherent style, further confirming our relative control.

\begin{table*}[ht]
\centering
\small
\resizebox{\textwidth}{!}{
\setlength{\tabcolsep}{4pt}
\begin{tabular}{lccccccc}
\toprule
\textbf{Ref $\backslash$ Target} & \textbf{Angry} & \textbf{Disgusted} & \textbf{Fear} & \textbf{Happy} & \textbf{Sad} & \textbf{Surprised} & \textbf{Neutral} \\
\midrule
\textbf{Angry} & -- & 48.4/63.4/78.6/\textbf{82.4} & 61.2/71.2/80.3/\textbf{88.7} & 78.4/84.2/86.2/\textbf{92.1} & 65.8/72.1/79.4/\textbf{86.8} & 64.7/74.5/80.7/\textbf{90.3} & 58.4/74.2/78.9/\textbf{84.6} \\
\textbf{Disgusted} & 61.9/70.2/80.2/\textbf{87.1} & -- & 57.8/70.6/80.4/\textbf{89.9} & 83.5/86.5/89.3/\textbf{96.8} & 73.5/78.3/80.6/\textbf{85.6} & 59.8/75.2/81.3/\textbf{92.7} & 61.9/73.9/80.5/\textbf{88.2} \\
\textbf{Fear} & 51.4/60.9/79.2/\textbf{85.7} & 65.7/75.9/78.5/\textbf{83.5} & -- & 82.9/85.7/90.4/\textbf{100.0} & 80.0/84.3/86.7/\textbf{92.9} & 55.7/73.4/81.2/\textbf{95.1} & 54.3/68.2/78.6/\textbf{81.7} \\
\textbf{Happy} & 65.2/73.5/88.5/\textbf{100.0} & 55.8/65.2/79.3/\textbf{86.2} & 52.1/70.2/80.6/\textbf{90.8} & -- & 81.1/82.0/81.4/\textbf{83.9} & 62.6/70.5/80.2/\textbf{87.6} & 52.3/65.4/81.6/\textbf{91.4} \\
\textbf{Sad} & 55.4/65.2/78.8/\textbf{81.6} & 66.8/75.5/80.3/\textbf{88.0} & 67.2/75.2/80.5/\textbf{85.3} & 81.1/83.2/86.6/\textbf{97.4} & -- & 60.5/72.5/79.7/\textbf{83.8} & 57.1/72.3/80.4/\textbf{90.9} \\
\textbf{Surprised} & 72.0/78.5/83.6/\textbf{100.0} & 57.0/65.4/80.2/\textbf{90.4} & 58.9/70.0/79.5/\textbf{82.7} & 72.0/79.2/79.8/\textbf{84.0} & 62.0/78.9/82.3/\textbf{92.0} & -- & 62.0/73.6/80.7/\textbf{88.5} \\
\textbf{Neutral} & 53.3/72.9/80.6/\textbf{94.4} & 57.8/70.1/79.2/\textbf{85.6} & 64.2/80.2/83.7/\textbf{91.2} & 70.7/78.5/80.3/\textbf{88.9} & 66.7/75.4/78.4/\textbf{83.3} & 67.8/72.9/80.5/\textbf{86.9} & -- \\
\bottomrule
\end{tabular}
}
\caption{Emotion transfer matrix for contradictory-style generation. Each off-diagonal cell reports the ACC (\%) in the format \textbf{CosyVoice / EmoVoice / IndexTTS2 / ReStyle-TTS}.}
\label{tab:contradict_emotion_matrix}
\end{table*}

\begin{table}[ht]
\centering
\small
\resizebox{\linewidth}{!}{
\setlength{\tabcolsep}{5pt}
\begin{tabular}{lccc}
\toprule
\textbf{Ref $\rightarrow$ Target} & \textbf{CosyVoice} & \textbf{EmoVoice} & \textbf{ReStyle-TTS} \\
\midrule
\multicolumn{4}{l}{\textbf{Pitch}} \\
Low $\rightarrow$ High & 74.9 & 72.4 & \textbf{90.2}  \\
High $\rightarrow$ Low & 76.9 & 73.1 & \textbf{92.8} \\
\midrule
\multicolumn{4}{l}{\textbf{Energy}} \\
Low $\rightarrow$ High & 87.5 & 76.1 & \textbf{92.4} \\
High $\rightarrow$ Low & 88.6 & 75.9 & \textbf{93.0} \\
\bottomrule
\end{tabular}
}
\caption{
Contradictory-style generation results for pitch and energy.
}
\label{tab:contradict_prosody_energy}
\end{table}

\subsection{Contradictory-Style Generation}
We further evaluate ReStyle-TTS in a contradictory-style setting on the VccmDataset test set, where the reference audio and the target style intentionally do not match. Concretely, the reference provides the speaker's timbre but carries an emotion or prosodic pattern that is different from the desired target. A more detailed explanation is provided in Appendix \ref{appendix:contradictory}.

Table~\ref{tab:contradict_emotion_matrix} reports the results for emotion transfer under such mismatched conditions. Compared with text-prompt-based controllable TTS, ReStyle-TTS more reliably follows the target emotion instead of the emotion implied by the reference audio, indicating that weakening the reference-style dependence via DCFG and then applying Style-LoRAs is effective for overriding the original style. We also examine contradictory-style control over pitch and energy in Table~\ref{tab:contradict_prosody_energy}. The results show that our method can consistently move pitch and energy in the desired opposite direction. We also provide a subjective evaluation of MOS-SA in Appendix \ref{appendix:contradictory}. These experiments confirm that ReStyle-TTS can handle challenging contradictory-style generation scenarios for both emotion and prosody.

\subsection{Ablation Studies}
We conducted ablation studies on DCFG and TCO in Table \ref{tab:ablation_dcfg_tco} and provide additional ablation studies on Orthogonal LoRA Fusion and the hyperparameter selection of $\lambda_a$ in DCFG in Appendix \ref{appendix:ablation}. The reported metrics in Table \ref{tab:ablation_dcfg_tco} are averaged over the 10 attributes shown in Figure \ref{fig:single-attr}, with LoRA strengths set to 2.0 for prosody control and 4.0 for emotion control. Regarding the control intensity metric $\text{Attr } \Delta \text{ (rel.)}$, we calculate the relative percentage change for prosody attributes and the absolute change in logits for emotion attributes, ensuring that the magnitudes remain comparable across different attribute types.

With standard CFG, text and reference guidance are coupled, preventing independent control. A high CFG weight (e.g., $\lambda_{\text{cfg}}=2$, equivalent to $\lambda_t=2$ and $\lambda_a=3$) enforces strong text fidelity and speaker similarity but severely limits style controllability. Conversely, a low CFG weight (e.g., $\lambda_{\text{cfg}}=-0.5$, equivalent to $\lambda_t=-0.5$ and $\lambda_a=0.5$) leads to severe distortion and unusable WER exceeding $1.0$, which is omitted in the Table. An intermediate setting ($\lambda_{\text{cfg}}=0.5$, i.e., $\lambda_t=0.5$ and $\lambda_a=1.5$) maintains intelligibility but still relies too heavily on the reference to enable effective style control. 
Overall, under CFG, improving controllability inevitably degrades text fidelity, and there exists no suitable value that can simultaneously achieve both controllability and text faithfulness. This motivates DCFG, which decouples and independently calibrates text and reference guidance. Furthermore, removing the Timbre Consistency Optimization module leads to a marked decline in speaker similarity, demonstrating its critical role in preserving timbre when reference guidance is reduced. 

\begin{table}[t]
\centering
\resizebox{\linewidth}{!}{
\begin{tabular}{lccc}
\toprule
\textbf{Setting} 
& \textbf{Attr $\Delta$ (rel.)} $\uparrow$ 
& \textbf{WER(\%)} $\downarrow$ 
& \textbf{Spk-sv} $\uparrow$ \\
\midrule
default ($\lambda_{\text{t}} = 2, \lambda_{\text{a}} = 0.5$) & 51.2\% & 2.31 & 0.79 \\
w/o DCFG ($\lambda_{\text{cfg}} = 2$)    
& 2.1\% & 1.83 & 0.90 \\
w/o DCFG ($\lambda_{\text{cfg}} = 0.5$)  
& 7.6\% & 2.67 & 0.85 \\
w/o TCO                                    
& 51.0\% & 2.32 & 0.71 \\
\bottomrule
\end{tabular}
}
\caption{Ablation study on DCFG and TCO.}
\label{tab:ablation_dcfg_tco}
\end{table}

\section{Conclusion}
In this paper, we propose ReStyle-TTS designed to enable continuous and relative style control in zero-shot speech synthesis. To achieve this, we first introduced Decoupled Classifier-Free Guidance (DCFG) to relax reference audio dependency while maintaining text fidelity. To achieve flexible manipulation, we leveraged Style-LoRAs with Orthogonal LoRA Fusion, allowing for the precise, simultaneous adjustment of multiple attributes. Furthermore, Timbre Consistency Optimization (TCO) was incorporated to ensure robust identity preservation. Experiments demonstrate that ReStyle-TTS effectively supports user-friendly style control and excels in challenging contradictory-style generation scenarios, offering a practical solution for expressive and controllable speech synthesis.

\clearpage

\section*{Limitations}
Although ReStyle-TTS successfully enables user-friendly relative and continuous style control, a primary limitation lies in its scalability to new attributes. Specifically, introducing control for a new style dimension requires collecting a corresponding dataset and performing additional LoRA fine-tuning. 

\section*{Ethics and Potential Risks}
The advancement of high-fidelity, zero-shot text-to-speech systems with flexible style control, such as ReStyle-TTS, brings significant ethical considerations. While our framework enhances user interaction and creative content generation, the ability to clone a speaker's identity from a short audio clip and manipulate their emotional expression poses potential risks for misuse, such as unauthorized voice cloning, deepfake generation, and the spread of misinformation. To mitigate these risks, it is crucial to ensure that such models are deployed responsibly. We strongly advocate for the integration of robust audio watermarking, the continuous development of synthetic speech detection models, and the establishment of strict protocols requiring explicit consent from voice contributors prior to synthesis.

\bibliography{custom}

\appendix
\section{Equivalence Between DCFG and Standard CFG}
\label{appendix:dcfg_equivalence}

We show that standard classifier-free guidance (CFG) is a special case of our Decoupled CFG (DCFG).

\[
\begin{aligned}
\hat{f}_{\text{CFG}}
&= f_{a,t} + \lambda_{\text{cfg}} (f_{a,t} - f_{\varnothing,\varnothing}) \\[3pt]
&= (1+\lambda_{\text{cfg}})\, f_{a,t} - \lambda_{\text{cfg}}\, f_{\varnothing,\varnothing}.
\end{aligned}
\tag{1}
\]

\[
\begin{aligned}
\hat{f}_{\text{DCFG}}
&= f_{\varnothing,t}
+ \lambda_t (f_{\varnothing,t} - f_{\varnothing,\varnothing})
+ \lambda_a (f_{a,t} - f_{\varnothing,t}) \\
&= (1+\lambda_t - \lambda_a)\, f_{\varnothing,t}
+ \lambda_a\, f_{a,t}
- \lambda_t\, f_{\varnothing,\varnothing}.
\end{aligned}
\tag{2}
\]

To recover the CFG form, the coefficients of the three terms must match between (1) and (2):
\[
\begin{cases}
1 + \lambda_t - \lambda_a = 0, \\
\lambda_a = 1 + \lambda_{\text{cfg}}, \\
\lambda_t = \lambda_{\text{cfg}}.
\end{cases}
\]

Solving yields:
\[
\lambda_a = 1 + \lambda_{\text{cfg}}, \qquad
\lambda_t = \lambda_{\text{cfg}}.
\]

Substituting these values into (2) gives exactly the standard CFG expression.

\section{Evaluation Details}
\label{appendix:eval}
For objective evaluations, following ControlSpeech~\citep{ji2024controlspeech}, we report not only the Word Error Rate (WER) and timbre similarity (Spk-sv) between the reference and synthesized speech, but also measure attribute-specific control effectiveness. For WER, we employ Whisper-large-v3~\citep{radford2023robust} for transcription. To evaluate timbre similarity (Spk-sv) between the original prompt and the synthesized speech, we utilize the base-plus-sv version of WavLM~\citep{chen2022wavlm}. For volume, we compute the $\ell_2$ norm of the amplitude of each short-time Fourier transform (STFT) frame. Pitch values are estimated using the Parselmouth toolkit, which extracts the fundamental frequency ($f_0$) and computes the geometric mean across all voiced regions. To evaluate emotion, we employ the official Emotion2Vec model \citep{ma2023emotion2vec} to compute speech emotion logits and classification accuracy. For subjective evaluations, we conduct MOS-SA (Mean Opinion Score – Style Accuracy) evaluations to measure the accuracy of the synthesized speech’s style via crowdsourcing. We randomly select 30 samples from the test set for subjective evaluation, and each audio sample is listened to by at least 10 testers. Testers are asked to rate the style accuracy on a 5-point scale ranging from 1 to 5.

\begin{table*}[t]
\centering
\small
\resizebox{\textwidth}{!}{
\setlength{\tabcolsep}{4pt}
\begin{tabular}{lccccccc}
\toprule
\textbf{Ref $\backslash$ Target} & \textbf{Angry} & \textbf{Disgusted} & \textbf{Fear} & \textbf{Happy} & \textbf{Sad} & \textbf{Surprised} & \textbf{Neutral} \\
\midrule
\textbf{Angry} & -- & 3.42/3.61/3.95/\textbf{4.18} & 3.58/3.79/4.05/\textbf{4.31} & 3.91/4.12/4.32/\textbf{4.52} & 3.49/3.72/4.00/\textbf{4.23} & 3.63/3.84/4.15/\textbf{4.41} & 3.38/3.68/3.95/\textbf{4.09} \\
\textbf{Disgusted} & 3.61/3.82/4.05/\textbf{4.29} & -- & 3.47/3.76/4.10/\textbf{4.36} & 4.02/4.21/4.40/\textbf{4.61} & 3.71/3.93/4.10/\textbf{4.24} & 3.46/3.81/4.15/\textbf{4.47} & 3.59/3.74/4.05/\textbf{4.28} \\
\textbf{Fear} & 3.33/3.54/3.95/\textbf{4.17} & 3.68/3.91/4.00/\textbf{4.08} & -- & 3.98/4.23/4.45/\textbf{4.68} & 3.87/4.06/4.25/\textbf{4.39} & 3.41/3.69/4.10/\textbf{4.46} & 3.29/3.57/3.95/\textbf{4.02} \\
\textbf{Happy} & 3.74/3.92/4.35/\textbf{4.69} & 3.36/3.58/3.95/\textbf{4.21} & 3.31/3.69/4.05/\textbf{4.43} & -- & 4.01/4.12/4.20/\textbf{4.29} & 3.62/3.83/4.10/\textbf{4.34} & 3.35/3.62/4.10/\textbf{4.42} \\
\textbf{Sad} & 3.45/3.64/3.95/\textbf{4.13} & 3.72/3.91/4.10/\textbf{4.31} & 3.63/3.82/4.05/\textbf{4.25} & 4.03/4.14/4.35/\textbf{4.58} & -- & 3.52/3.79/4.00/\textbf{4.16} & 3.44/3.71/4.10/\textbf{4.37} \\
\textbf{Surprised} & 3.88/4.09/4.40/\textbf{4.71} & 3.41/3.63/4.00/\textbf{4.27} & 3.49/3.71/3.95/\textbf{4.12} & 3.91/4.03/4.20/\textbf{4.35} & 3.58/3.97/4.25/\textbf{4.49} & -- & 3.61/3.83/4.10/\textbf{4.32} \\
\textbf{Neutral} & 3.39/3.78/4.20/\textbf{4.46} & 3.51/3.69/4.00/\textbf{4.24} & 3.67/3.99/4.20/\textbf{4.41} & 3.79/4.01/4.20/\textbf{4.43} & 3.55/3.77/4.00/\textbf{4.14} & 3.66/3.81/4.10/\textbf{4.34} & -- \\
\bottomrule
\end{tabular}
}
\caption{Emotion transfer matrix for contradictory-style generation. Each off-diagonal cell reports the MOS-SA (5-point scale) in the format \textbf{CosyVoice / EmoVoice / IndexTTS2 / ReStyle-TTS}.}
\label{tab:contradict_emotion_matrix_mos}
\end{table*}

\section{Relative Style Control }
\label{appendix:relative}

\begin{figure}[h]
    \centering
    \includegraphics[width=0.8\linewidth]{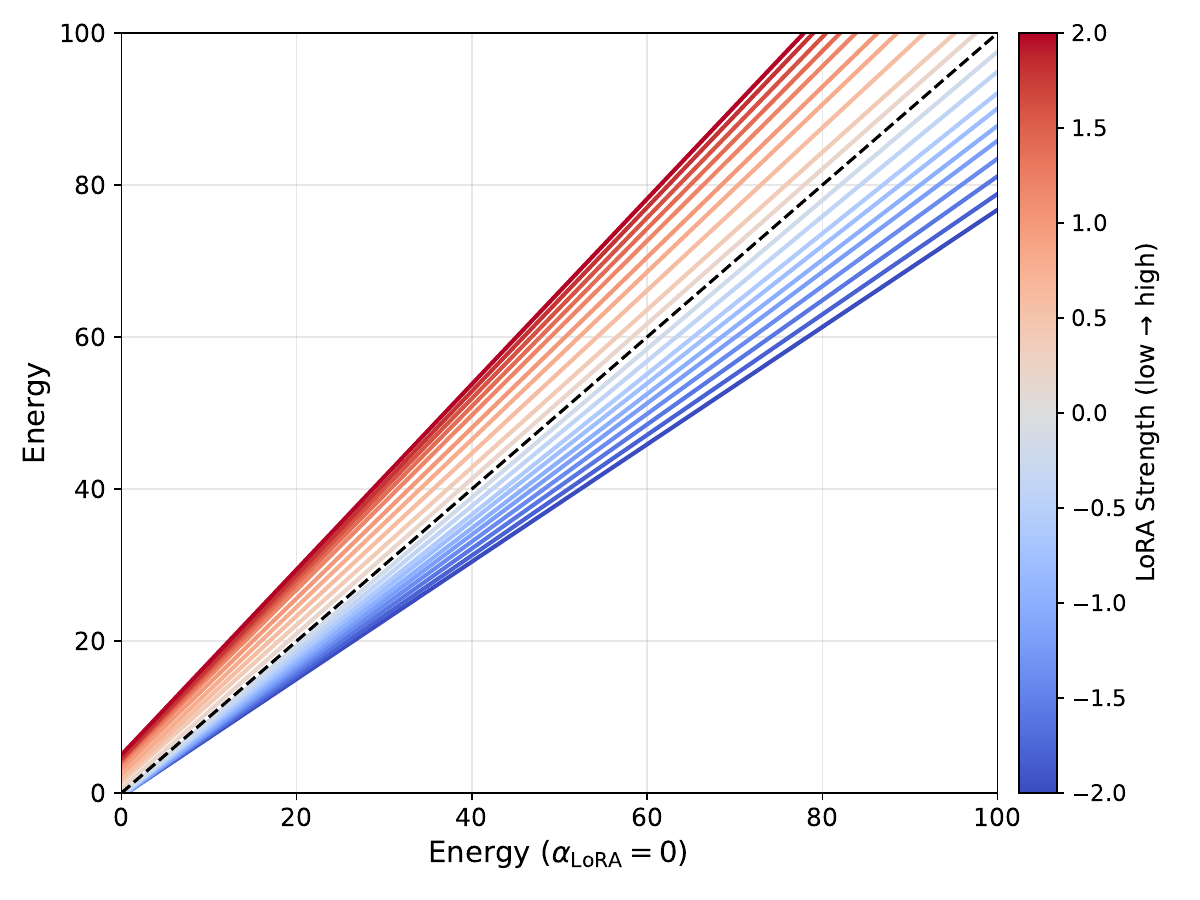}
    \caption{
    Linear regression analysis of energy control across different LoRA scales.
    }
    \label{fig:relative_high_energy_fit_lines}
\end{figure}

\begin{figure}[h]
    \centering
    \includegraphics[width=0.8\linewidth]{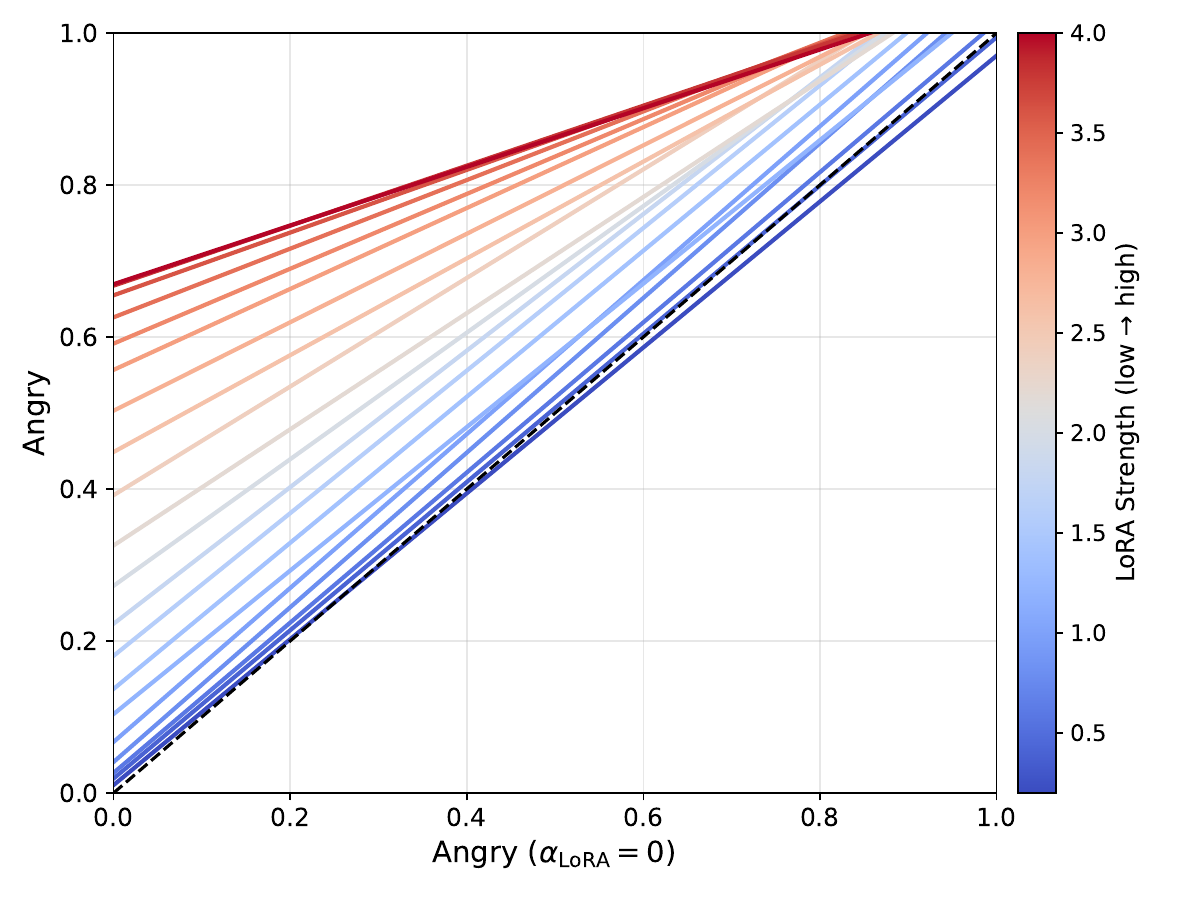}
    \caption{
    Linear regression analysis of angry control across different LoRA scales.
    }
    \label{fig:relative_angry_fit_lines}
\end{figure}

As illustrated in Figure \ref{fig:relative_high_energy_fit_lines}, we visualize the relationship between the baseline style and the modified style. The x-axis represents the energy of the generated speech when the LoRA scale is set to 0 (serving as the reference baseline), while the y-axis displays the energy values obtained under varying LoRA strengths. 

It can be observed that as the LoRA strength increases, the slope of the fitted regression line steepens, rising from 0.77 to 1.22. This monotonic increase in slope demonstrates that ReStyle-TTS achieves true relative control by scaling the inherent attributes of the reference audio rather than overwriting them with fixed absolute values. In addition to relative prosody control, we also present the results for angry in Figure \ref{fig:relative_angry_fit_lines} as an example of relative emotion control.

\section{Contradictory-Style Generation}
\label{appendix:contradictory}
In this section, we provide a detailed explanation of the experimental setup for Contradictory-Style Generation. Our approach diverges from the original usage of the VccmDataset in ControlSpeech. Each sample in the VccmDataset consists of an audio clip, its corresponding transcription, and a style prompt. However, the original ControlSpeech evaluation did not address scenarios where the style of the reference audio conflicts with the target style. For instance, attempting to synthesize angry speech when only a happy reference clip is available. To evaluate this capability, we utilize the audio samples from the VccmDataset as reference audio. For each reference, we attempt to synthesize speech targeting every emotion category that differs from the reference's inherent emotion. The synthesis accuracy is then quantitatively evaluated using the Emotion2Vec model. Regarding the specific control configurations: for ReStyle-TTS, we simply apply the Style-LoRA corresponding to the target emotion. For the natural language-controlled baselines, CosyVoice and EmoVoice, we provide the specific style instruction: `I'm saying this with great \{emotion\}.' We also provide a subjective evaluation of MOS-SA in Table \ref{tab:contradict_emotion_matrix_mos}. The results are consistent with the objective evaluation, and our ReStyle-TTS achieves the best performance across all contradictory-style generation scenarios.

\section{Ablation Studies}
\label{appendix:ablation}
\begin{figure}[t]
    \centering
    \includegraphics[width=1\linewidth]{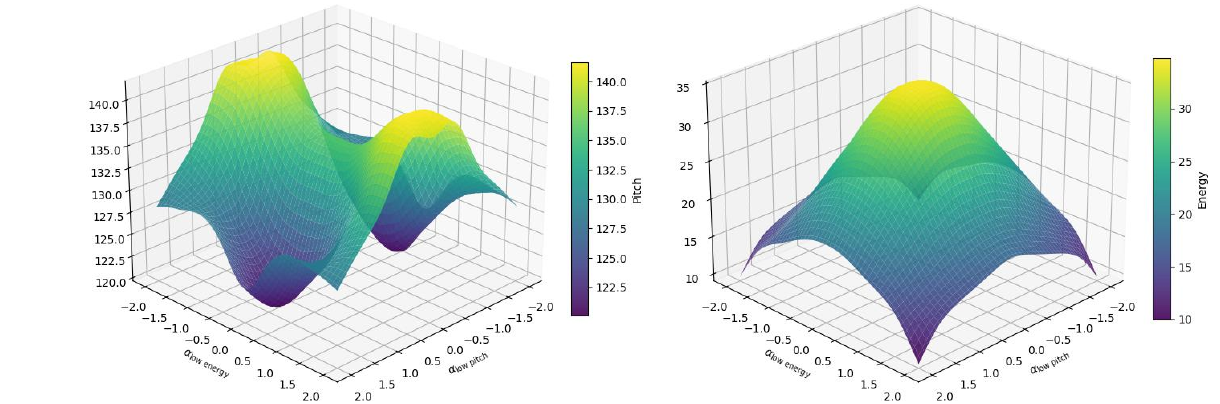}
    \caption{Ablation Study of Orthogonal LoRA Fusion.}
    \label{fig:ablation_fusion}
\end{figure}

For Orthogonal LoRA Fusion, we also conducted ablation studies. As shown in Figure~\ref{fig:ablation_fusion}, the control of pitch and energy becomes fully entangled, making it impossible to adjust them independently. 

\begin{figure}[t]
    \centering
    \includegraphics[width=0.8\linewidth]{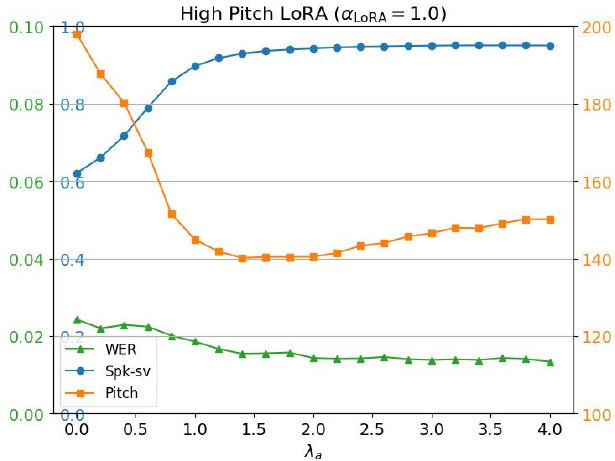}
    \caption{Ablation Study of $\lambda_a$.}
    \label{fig:ablation_lamda_a}
\end{figure}

We further performed ablation experiments on the parameter $\lambda_a$. Since DCFG decouples the model’s reliance on the text and the reference audio, we fixed $\lambda_t = 2$, a commonly used setting to maintain strong text dependency, and varied $\lambda_a$, which governs the trade-off between timbre similarity and controllability. The results are shown in Figure~\ref{fig:ablation_lamda_a}. It can be observed that $\lambda_a$ governs the trade-off between timbre similarity and controllability. We ultimately chose $\lambda_a = 0.5$, sacrificing some timbre similarity to achieve better controllability, while using Timbre Consistency Optimization to compensate for the lost timbre similarity.

\begin{table*}[t]
\centering
\small
\caption{Single-attribute control results using CosyVoice backbone.}
\label{tab:cosy_single}
\begin{tabular}{lccccccccc}
\toprule
\textbf{Strength} & \textbf{-2.0} & \textbf{-1.5} & \textbf{-1.0} & \textbf{-0.5} & \textbf{0.0} & \textbf{+0.5} & \textbf{+1.0} & \textbf{+1.5} & \textbf{+2.0} \\
\midrule
Pitch & 165.1 & 160.8 & 155.4 & 148.9 & 142.5 & 136.2 & 130.8 & 125.6 & 122.7 \\
Energy & 49.8 & 47.2 & 43.5 & 39.1 & 35.8 & 31.4 & 28.2 & 25.1 & 23.6 \\
\midrule
\textbf{Strength} & \textbf{0.0} & \textbf{0.5} & \textbf{1.0} & \textbf{1.5} & \textbf{2.0} & \textbf{2.5} & \textbf{3.0} & \textbf{3.5} & \textbf{4.0} \\
\midrule
Angry & 0.10 & 0.16 & 0.28 & 0.42 & 0.56 & 0.68 & 0.74 & 0.77 & 0.79 \\
Happy & 0.06 & 0.12 & 0.24 & 0.38 & 0.52 & 0.64 & 0.71 & 0.75 & 0.76 \\
\bottomrule
\end{tabular}
\end{table*}

\section{Generalization to Different Model Backbones}
\label{appendix:cosyvoice_generalization}

We further verify the generalizability of our proposed framework by applying it to a different model backbone, CosyVoice \citep{du2024cosyvoice2}. For rapid validation, we evaluated Low Pitch, Low Energy, Angry, and Happy attributes. Notably, we rely solely on LoRA strength to modulate the style of CosyVoice, bypassing its native text-based instructions. The results demonstrate that ReStyle-TTS generalizes effectively to different backbones for both single-attribute and multi-attribute control.

As shown in Table \ref{tab:cosy_single}, sweeping the LoRA strength on the CosyVoice backbone yields smooth and monotonic transitions for both prosodic and emotional attributes, consistent with the observations on F5-TTS. Furthermore, we applied Orthogonal LoRA Fusion (OLoRA) to CosyVoice for joint Energy-Pitch control. As illustrated in Table \ref{tab:cosy_joint}, adjusting the strength of one attribute has minimal impact on the other (e.g., varying Pitch strength across a 30Hz range results in less than 1.0 unit of Energy fluctuation). These results confirm that our decoupling and fusion mechanism is backbone-agnostic and maintains high disentanglement across different model architectures.

\begin{table}[h]
\centering
\small
\caption{Joint Energy-Pitch control on CosyVoice using OLoRA.}
\label{tab:cosy_joint}
\begin{tabular}{lccccc}
\toprule
\textbf{Energy \textbackslash{} Pitch} & \textbf{-2.0} & \textbf{-1.0} & \textbf{0.0} & \textbf{+1.0} & \textbf{+2.0} \\
\midrule
\multicolumn{6}{c}{\textit{Measured Pitch}} \\
\textbf{-2.0} & 147.0 & 142.5 & 135.5 & 128.5 & 121.5 \\
\textbf{-1.0} & 147.8 & 142.7 & 135.7 & 128.7 & 121.7 \\
\textbf{0.0}  & 148.5 & 143.0 & 136.0 & 129.0 & 122.0 \\
\textbf{+1.0} & 149.3 & 143.3 & 136.3 & 129.3 & 122.3 \\
\textbf{+2.0} & 150.0 & 143.5 & 136.5 & 129.5 & 122.5 \\
\midrule
\multicolumn{6}{c}{\textit{Measured Energy}} \\
\textbf{-2.0} & 42.5 & 42.2 & 42.0 & 41.8 & 41.5 \\
\textbf{-1.0} & 38.5 & 38.2 & 38.0 & 37.8 & 37.5 \\
\textbf{0.0}  & 34.5 & 34.2 & 34.0 & 33.8 & 33.5 \\
\textbf{+1.0} & 30.5 & 30.2 & 30.0 & 29.8 & 29.5 \\
\textbf{+2.0} & 26.5 & 26.2 & 26.0 & 25.8 & 25.5 \\
\bottomrule
\end{tabular}
\end{table}

\end{document}